# Anomalous and Planar Nernst effects in thin-films of half-metallic ferromagnet $La_{2/3}Sr_{1/3}MnO_3$

*Cong Tinh Bui, and Francisco Rivadulla\**

\*e-mail: f.rivadulla@usc.es

Centro de Investigación en Química Biológica y Materiales Moleculares (CIQUS), Universidad de Santiago de Compostela, 15782-Santiago de Compostela, Spain.

We report the planar and anomalous Nernst effect in epitaxial thin films of spin polarized $La_{2/3}Sr_{1/3}MnO_3$. The thermal counterpart of the anomalous Hall effect in this material (i.e. the anomalous Nernst effect) shows a extreme sensitivity to any parasitic thermal gradient, resulting in large asymmetric voltages under small temperature differences. This should be considered when interpreting the magnitude of the electrical response in nanostructures and devices that operate under high current densities. Finally, none of the observed magneto-thermoelectric signals is compatible with the observation of the Spin Seebeck Effect in this material.

The discovery of intrinsic Spin Seebeck Effect (SSE) in magnetic materials, irrespective of their conductivity, opens unforeseen possibilities for the creation and manipulation of pure spin currents (spin caloritronics) as well as of energy harvesting [1,2,3]. Basically, when a thermal gradient is established parallel to the magnetization of a ferromagnet (FM), spin angular momentum is transported along the system in response to the temperature difference. This time-varying magnetization is able to pump a pure spin current at the interface with a paramagnetic metal (normally Pt), which is then transformed into a transverse electrical current via inverse spin Hall effect [4].

In FM metals, the different density of states and Fermi velocities for the spin up/down population produce different conductivities for the opposite spin directions [5]. Therefore when the spin lifetime is larger than the momentum relaxation time, a *spin-* *dependent* Seebeck and Peltier coefficient are predicted, on the basis of the Onsager reciprocities [6].

However, the observation of a clear SSE signal in magnetic insulators demonstrates that its origin must be different to the conventional Seebeck effect in magnetic conductors, though their phenomenology is very similar [2]. Although the role of spin-phonon coupling through the substrate was emphasized to explain its long-range nature [7,8], the microscopic mechanism is still under debate. On the other hand, recent studies on metallic FM Permaloy have shown that the anomalous Nernst effect (ANE) may be an important contribution to the measured SSE in some materials [9,10,11].

Therefore, it is of fundamental interest to explore the nature and magnitude of spin-dependent thermoelectric effects in novel technological materials, and thereby provide a better understanding of the delicate balance between spin, charge and heat currents in nanodevices [12,13,14,15,16].

In this paper we report the observation of an intrinsic planar Nernst effect (PNE) and anomalous Nernst effect (ANE) in thin films of $La_{2/3}Sr_{1/3}MnO_3$ (LSMO). This material shows a fully spin polarized *3d* band [17], and a $T_C \approx 360$ K, which motivated its extended use as a FM electrode in tunnel junctions [18]. We show that through a careful control of the thermal gradients the ANE can be separated from the symmetric PNE response. We further demonstrate that there is a perfect correspondence between the magnetothermal effects and their electrical counterparts in LSMO. Our findings also establish an upper limit for the possible observation of SSE in this system. These new findings are relevant for a better understanding of the spin-dependent thermoelectric phenomena in similar correlated metallic oxides.

Before discussing the results, we briefly recall the physical quantities that govern the magneto-thermoelectric effects in our system. In magnetic conductors, the spin-orbit interaction introduces an anisotropic thermoelectric voltage depending on the angle, $\Theta$, between the temperature gradient and the magnetization, $M$ [6]. These are the thermal counterparts (Onsager reciprocals) of the anisotropic magnetoresistance (AMR) and planar Hall effect (PHE) [19]. For the PNE, the transverse voltage $V_{xy}$ is related to $M$ and $\Theta$ by [10,20]:

$$S_{xy} = \frac{V_{xy}}{\nabla T_x} \propto |M|^2 \sin\Theta \cos\Theta \qquad (1)$$

with $M$ and $\nabla T$ lying both in the *xy*-plane.

However, in a conducting FM any $\nabla T_z \neq 0$ will create a measurable $V_{xy}$ response due to the anomalous Nernst effect (ANE) [11]:

$$\nabla V_{xy} = -S_{xx}\xi\left(\hat{m}\times\nabla T_z\right) \qquad (2)$$

where $S_{xx}$ is the linear Seebeck coefficient, $\hat{m}$ is the unit vector of the magnetization, and $\xi$ is the Nernst factor [12].

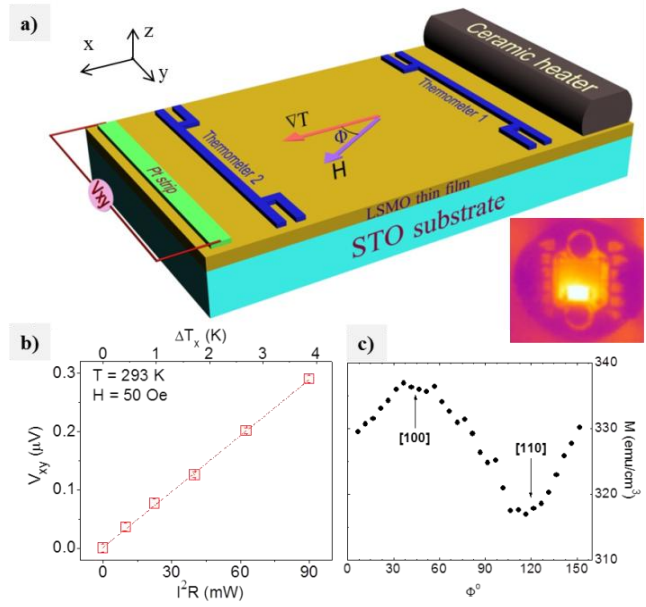

**Figure 1. a)** Sketch of the device used to measure the PNE and ANE in a thin film (35 nm thick) of LSMO, along with a thermal image of the actual device with $\nabla T_x \neq 0$. The two thermometers are Pt resistances deposited by optical lithography (see supporting information for details). **b)** Evolution of the transverse voltage with the heating power and $\nabla T_x$. **c)** Angular dependence of the magnetization ($H = 100$ Oe) showing the crystalline directions of the easy/hard axis in LSMO.

The experimental setup for the measurement of the magneto-thermoelectric effects is shown in Figure 1 (see supporting information for details). We have used the same transverse configuration as normally used to measure the SSE. A small $\nabla T_x \leq 0.8$ K/mm was always used in order to be within the linear, reversible regime for the thermopower, and to avoid any uncontrolled temperature gradient in other direction. In addition, we have also determined the transverse voltage for a cross-plane thermal gradient ($\nabla T_z \neq 0$) while keeping the in-plane temperature constant ($\nabla T_x = 0$).

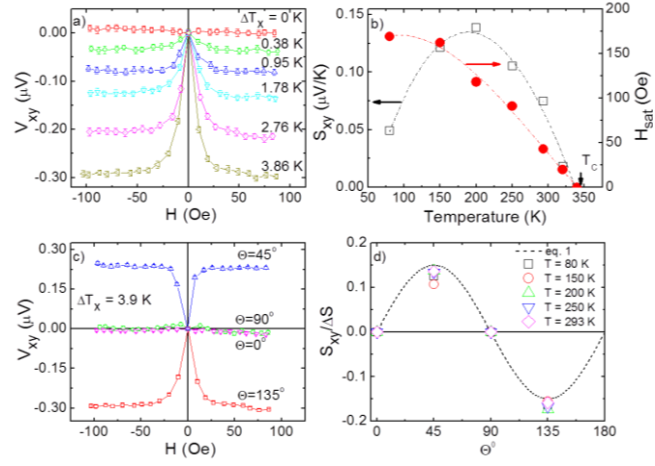

**Figure 2. a)** Field dependence of the transverse voltage, $V_{xy}$, at 293 K for different temperature gradients, $\nabla T_x$. **b)** Temperature dependence of the transverse thermoelectric power $S_{xy}$ and the saturation field. **c)** Transverse voltage at 293 K, for different angles $\Theta$. **d)** Angular dependence of the saturation transverse voltage ($V_{xy}$ at $H_{sat}$), normalized by its anisotropic response, according to Eq. (1), at different temperatures. The dashed line represents the $\cos\Theta\sin\Theta$ characteristic of PNE, multiplied by a temperature independent fitting factor. A constant thermoelectric offset voltage was subtracted in a) and c) to center the curves at zero voltage when $H = 0$.

The in-plane angular dependence of the magnetization of LSMO is show in Fig. 1(c). The result is the characteristic of a system with biaxial symmetry, with the easy axis along the [100] direction of the film [21]. The transverse voltage, $V_{xy}$, was recorded as a function of the magnetic field for different angles, $\Theta$, between $\nabla T_x$ and $M$ (according to the easy axis direction shown in Fig. 1c). The results are summarized in Figure 2: $V_{xy}$ decreases with the applied magnetic field, until saturation is reached at $H_{sat}$ (~ 43 Oe). The amplitude of $V_{xy}$ depends linearly on the temperature gradient [Fig. 1(b)]. Furthermore, all curves collapse when divided by the corresponding temperature gradient (see supporting information, Fig. S3), indicative of its thermoelectric origin, and demonstrates the accurate determination of $\nabla T_x$ by the two Pt-resistances. Finally, the appearance of symmetric behavior in all curves indicates that the PNE is certainly the driving mechanism of the observed effect [10].

The amplitude of the magnetothermoelectric power $S_{xy}$ and $H_{sat}$ both drop to zero when base temperature approaches $T_C$ [Fig. 2(b)] demonstrating their intimate relationship to the spontaneous magnetization.

The results of $V_{xy}$ at 293 K (with $\Delta T_x = 3.9$ K) when rotating the magnetic field are show in Fig. 2(c) and the normalized magnetothermopower, $S_{xy}/\Delta S$, is shown in Fig. 2(d), for different base temperatures. The angular dependence of $S_{xy}$ shows a good agreement with the predictions of Eq. (1) for PNE, at all the temperatures probed in this work [Fig. 2(d)].

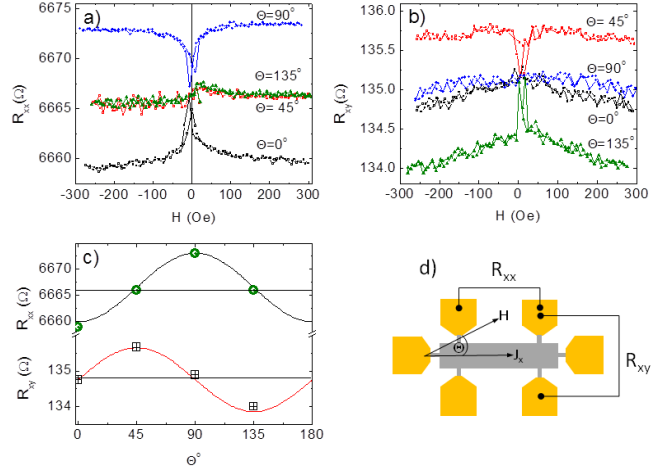

**Figure 3. a)** Longitudinal ($R_{xx}$, AMR) and **b)** transverse ($R_{xy}$, PHE) components of the magnetoresistance at 293 K, along with their angular dependence **(c)**. The sketch in **(d)** shows the relative orientation of the field and current during the experiment. The dimensions of the film channel in the Hall bar are $100 \times 500$ $\mu m^2$. The long axis of the Hall bar is along the (110) direction of LSMO film.

The PNE has its origin in the spin-orbit interaction, and should therefore present a perfect correspondence with the PHE in the same material [20]. In order to verify such a correspondence in the LSMO, we have measured the AMR ($R_{xx}$) and PHE ($R_{xy}$) in thin films grown under the same conditions as those described above. The results are shown in Figure 3. Both AMR and PHE are observed in LSMO at room temperature ($T << T_C$) [Fig. 3(a) and Fig. 3(b), respectively]. Moreover, the angular dependence of the PHE shown in Fig. 3(c) is similar to its thermal counterpart (PNE), confirming their common physical origin.

In addition, the maximum AMR should be shifted by 45º with respect to the PHE, according to [22]:

$$R_{xx} = \rho_\perp + (\rho_{ll} - \rho_\perp)\cos^2\Theta \qquad (3)$$

This was also verified experimentally in our LSMO film samples, as shown in Fig. 3(c).

We have also observed a non-saturating negative MR, at least up to 1.2 T, the maximum field probed in this work, which is already visible in Fig. 3(b). This is commonly observed in thin films of *3d* metals, due to the damping of spin waves at high fields which leads to the reduction of electron-magnon scattering [23]. The same phenomenon was observed in the thermal PNE (supporting information, Fig. S4).

Importantly, we have noticed that even a very small fluctuation (< 0.01 K) in the base temperature of the cryostat is enough to produce a small thermal gradient across the plane of the film ($\nabla T_z \neq 0$), which introduces an asymmetric response of the transverse voltage with respect to the magnetic field [Fig. 4(a)]. This asymmetric signal shows in addition to the contribution from the symmetric PNE, an odd function of the magnetic field.

As we mentioned before, in a conducing FM any $\nabla T_z \neq 0$ will create a measurable $V_{xy}$ response due to ANE (Eq. 2). This signal should change sign with the magnetic field, and therefore should be responsible for the asymmetric component observed in Figure 4.

In order to quantify the ANE response in LSMO, a small temperature gradient was intentionally applied perpendicular to the film surface ($\nabla T_z \neq 0$). The temperature of the Cu block under the film was varied by less than one Kelvin, while keeping $\nabla T_x = 0$. Given the thickness of the film (~ 35 nm) it is difficult to estimate an accurate value of the cross plane temperature difference. As shown in Fig. 4(b), a large transverse voltage is observed, in spite of the small temperature difference. The signal is an odd function of the magnetic field, following perfectly the magnetization of the sample, as expected for the ANE.

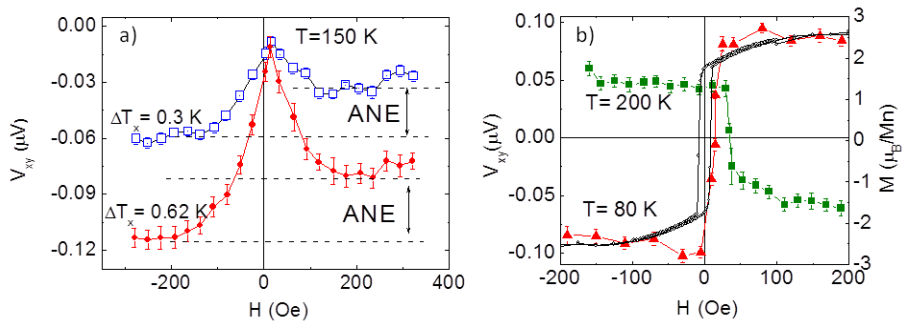

**Figure 4. a)** Mixed contribution of the odd ANE and even PNE field dependence, at 150 K, for different in plane thermal gradients, and $\nabla T_z \neq 0$. **b)** ANE at 200 K (squares) and 80 K (triangles), measured with $\nabla T_x = 0$ and $\nabla T_z \neq 0$. The solid line corresponds to the field dependence of the magnetization at 80 K, with the magnetic field applied along the film plane.

The temperature difference across the film can now be estimated from the experimental value of $V_{xy}$ and Eq. (2). At 200 K, the Seebeck coefficient of LSMO $S_{xx} = 2.3$ µV/K (supporting information, Fig. S5). Using $\xi = 0.13$, like for Permalloy [12], we obtain a $\Delta T_z \approx 2$ µK, which corresponds to a gradient of 0.6 K/cm across our 35 nm thick film of LSMO. Note that a substantial error in the estimation of $\xi$ will not change the conclusion, *i.e.* the extreme sensitivity of the transverse voltage to any parasitic cross-plane temperature gradient. This is an important result, because it demonstrates that very small parasitic thermal currents in magnetic conductors may produce thermoelectric voltages in the presence of a transverse magnetic field, which could be relevant in nanodevices.

Moreover, this transverse signal shows the same field and same angular dependence as the SSE, and therefore it must be carefully addressed in conducting ferromagnets. On the other hand, the analysis of the intrinsic magnetothermoelectric response in magnetic insulators can be complicated by the existence of magnetic moments induced at Pt by proximity effects close to the FM/Pt interface [24,25]. These results make it very difficult to separate the contribution from ANE and SSE, even in magnetic insulators, and independently of whether a transverse or longitudinal configuration is used to detect the SSE signal [26]. Therefore, other metals with smaller spin-orbit coupling than Pt, not so close to a magnetic instability, should probably be used instead to detect the intrinsic SSE [27]. In any case, our results put an upper limit to the possible observation of SSE in LSMO. For example, at 200K and with the conditions of our measurement the possible SSE signal, if there is any, should be much less than 50 nV [Fig. 4(b)].

Most important is our observation of a sign change of the ANE with temperature between 200 K and 80 K, even though the direction of the thermal gradient and magnetization is kept constant. In fact, comparing Fig. 4(a) and (b), it can be seen that the ANE already changes sign between 150K and 200K. This sign reversal of the ANE was observed in other systems, like semiconducting $Ga_{1-x}Mn_xAs$ [28], and Pt/YIG bilayers [24], and can be understood in terms of the Mott relationship between the anomalous Nernst and Hall coefficients [28,29,30,31]. This means that the ANE must be determined by the energy dependence of its electrical charge counterpart:

$$S_{xy} = \frac{-\pi^2 k_B^2 T}{3e} \left( \frac{\partial \ln \rho_{xy}}{\partial E} \right)_{E_F} \qquad (4)$$

Where $\rho_{xy} = \lambda \, M_z \, \rho_{xx}{}^n$ is the anomalous Hall resistivity [32,33]. Given the relationship between the anomalous Hall conductivity and the linear resistivity, $\sigma_{xy} \propto \rho_{xx}{}^{n-2}$, $n = 1$ results in a linear dependence of the Hall conductivity with the scattering time, $\tau$, characteristic of the skew scattering extrinsic mechanism for AHE in the clean (high conductivity) limit. When $\tau$ decreases, a transition towards the intrinsic regime ($n = 2$, $\sigma_{xy}$ independent of $\tau$) and the dirty regime ($n = 1.6$, hopping conductivity) occurs [29,34,35]. Transition metal-oxides with intermediate resistivities ($\rho_{xx}$ larger than $\approx 10$ $\mu\Omega$cm) are expected to fall in the intrinsic regime ($n = 2$) [29].

In the LSMO film, with an in-plane magnetization $M_x$ and a perpendicular temperature gradient $\nabla T_z \neq 0$, the anomalous Nernst signal of Eq. (2) can be expressed as a function of the other transport coefficients, and assuming $\sigma_{xy} = -\sigma_{yx}$, then [28]:

$$S_{xy} = \frac{\rho_{xy}}{\rho_{xx}}\left( \frac{-\pi^2 k_B^2 T}{3e} \frac{\lambda}{\lambda'} - (n-1)S_{xx} \right) \qquad (5)$$

where $\lambda' = (\partial \lambda / \partial E)$.

Given that the transverse and linear resistivities do not change sign between 150 K and 200 K [33], we can infer from the results in Fig. 4 that $n \neq 1$. This is in agreement with previous observations in similar oxides with comparable conductivities [29], and places LSMO in the intrinsic regime, as expected.

In summary, we have investigated the intrinsic magneto-thermoelectric coefficients of LSMO thin films and demonstrated experimentally the relationship among the different components of the transport tensors in the presence of $H$. Through a careful control of the thermal gradients in different directions of the samples we have disentangled the contributions of the PNE and ANE to the magnetothermal response of LSMO. Our results also established an upper limit for any possible observation of the SSE in this technologically important material.

The extreme sensitivity of ANE to $\nabla T_z$ could be useful to sense very small temperature differences across nanostructures, which are extremely difficult to measure accurately by other methods. Finally, our results also show the utility of ANE to study the mechanism of anomalous Hall effect in ferromagnetic metals.

**Acknowledgements**

This research was supported by the European Research Council (ERC StrG-259082, 2DTHERMS). We would like to thank Prof. Dr. M. Abd-Elmeguid, University of Cologne, Germany, for fruitful discussion.

# Supporting Information

### Anomalous and Planar Nernst effects in thin-films of half-metallic ferromagnet La$_{2/3}$Sr$_{1/3}$MnO$_3$

*Cong Tinh Bui, and Francisco Rivadulla\**

Epitaxial thin films of LSMO were grown on (001) TiO$_2$-terminated SrTiO$_3$ (STO) substrate by Pulsed Laser Deposition (F-Kr, 248 nm, 800 ºC, 200 mTorr O$_2$ pressure). The thickness determined by X-ray reflectivity is $\approx$ 35 nm (Fig. S1), which makes the transport properties similar to the bulk.

A ceramic heater ($R$ = 100 Ω, glued in good thermal contact to one end of the film) was used to create the temperature gradient along the $x$ direction, $\nabla T_x$, as shown in Fig. 1(a). To probe the transverse voltage $V_{xy}$, a Pt strip (4 mm × 100 μm × 8 nm) was deposited at the other end of the sample, right after thin film growth using a shadow mask, in order to achieve an optimum contact interface between Pt and LSMO. The transverse voltage was measured with a Keithley 2181 nanovoltmeter between the ends of the Pt bar. During measurements in magnetic field, the current was kept constant in the electromagnet at each field, in order to avoid parasitic currents in our measurement. More than 100 points were averaged at each field. The measurements were performed in high vacuum ($P < 10^{-6}$ mbar) to avoid any influence of convection.

The homogeneity of temperature in *y* direction and the distinctly established temperature gradient in *x* direction are clearly observed in thermal image of Fig. 1(a) when injecting current through the heater.

In order to probe accurately the temperature gradient across the sample, two metallic thermometers Cr/Pt (5nm/100nm, 100 µm wide) were deposited on the surface of the film, close to the positions of the heater and the Pt strip [Fig. 1(a) and Fig. S2]. Special care was taken to control the temperature gradients along the different directions of the sample. For this purpose the sample was mounted on top of a massive Cu block inside the cryostat, with a much larger area than the film, to avoid any uncontrolled variation in the temperature along or across the sample. Moreover, a small $\nabla T_x \leq 0.8$ K/mm was always used in order to be within the linear, reversible regime for the thermopower, and also to avoid any uncontrolled temperature gradient in other direction that could occur for larger temperature differences across the sample.

The X-ray diffraction pattern of the LSMO thin film sample around (002) peak of STO and LSMO and the X-ray reflectivity curve are shown in Figure S1. The thickness of the LSMO film can be precisely determined from the reflectivity curve.

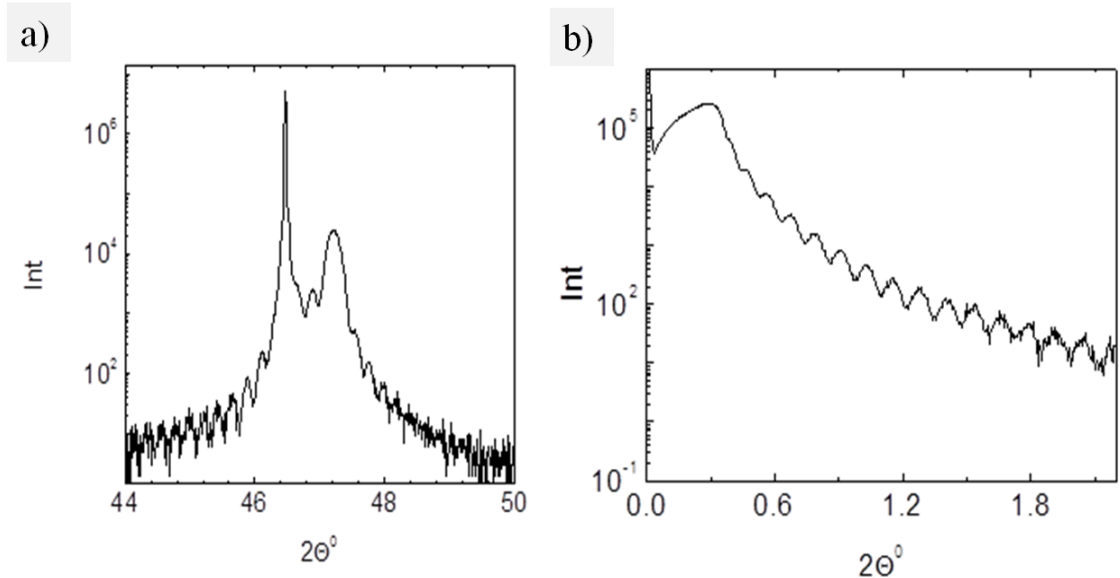

**Figure S1**: **a)** Detail of the X-ray diffraction pattern around the (002) peak of STO and LSMO. **b)** X-ray reflectivity curve of a LSMO film. The fitting gives a thickness of 35 nm.

The two thermometers consisting of 100nm thick Pt with 5nm thick Cr as adhesive layer (100 μm wide) were utilized to determine the temperature gradient between the heater and Pt metal strip. The temperature dependence of resistance of the two thermometers was calibrated in the temperature range of 77 K – 320 K. These calibration curves shown in Fig. S2(a) were then used to obtain the temperature at each thermometer based on 4-point resistance measurement. Fig. S2(b) shows the temperature gradient along $x$ direction, $\Delta T_x$, as a function of heating power passing through ceramic heater at different base temperatures.

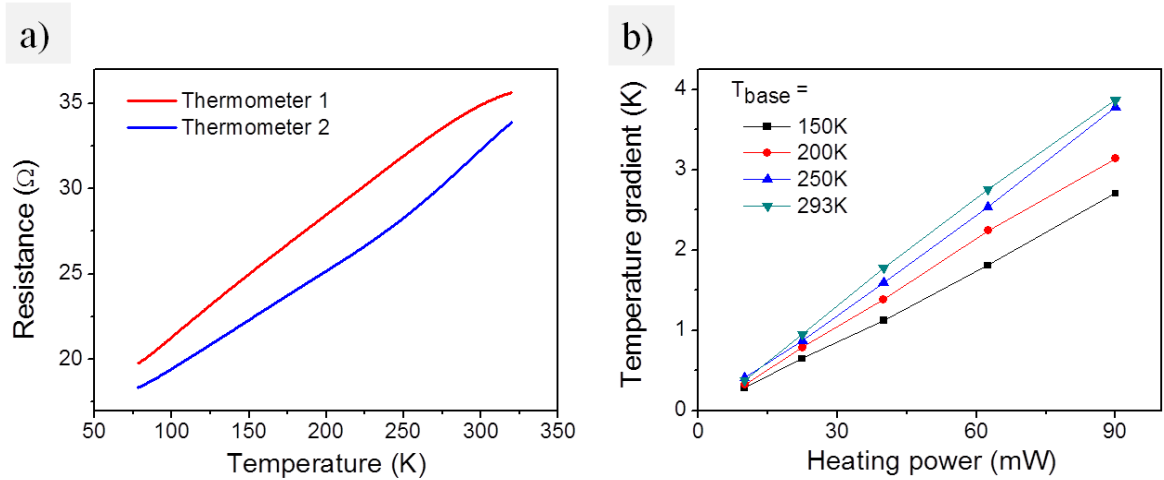

**Figure S2**: **a)** The temperature dependence of resistance of thermometer 1 & 2. **b)** The temperature gradient along $x$ direction, $\Delta T_x$, as a function of heating power at different base temperatures.

Figure S3 shows the normalization of the transverse voltage by the temperature gradient, corresponding to the data in Figure 2 of the paper. The coincidence of all curves indicates the linear response of the voltage with temperature gradient, consequently the thermoelectric origin of the observed effect.

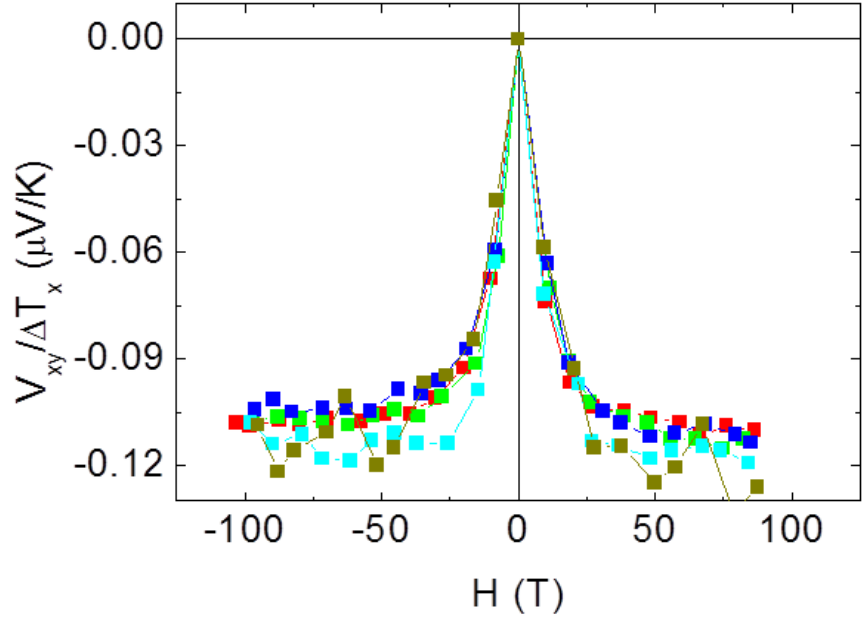

**Figure S3**: Transverse voltage $V_{xy}$ normalized to the thermal gradient ($\Delta T_x$), corresponding to the data presented in Fig. 2. The base temperature is 293 K.

The magnetoresistance (MR) and magneto-Seebeck of the LSMO sample measured in high magnetic field are shown in Fig. S4(a) and (b), respectively. The results indicate a non-saturating linear dependence at high field.

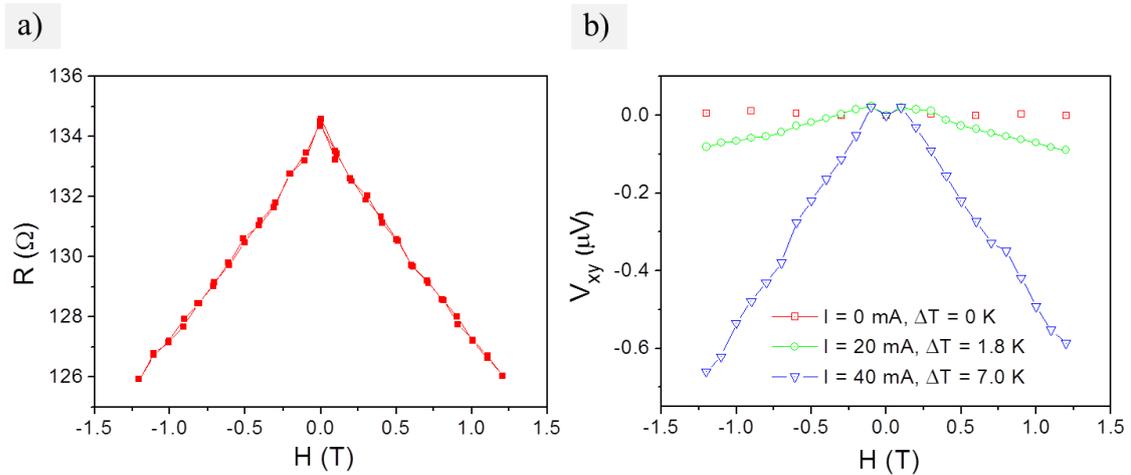

**Figure S4**: High field MR **(a)** and magneto-Seebeck **(b)**, showing the non-saturating linear dependence at high field.

Figure S5(a) displays the temperature dependence of the electrical resistivity of LSMO thin film. The Seebeck coefficient of a single crystal of LSMO (black) as well as of the 35 nm thin LSMO film sample (red) are shown in Fig. S5(b). A two-Pt thermometer configuration as described in the text was used to determine the Seebeck coefficient of the thin film sample.

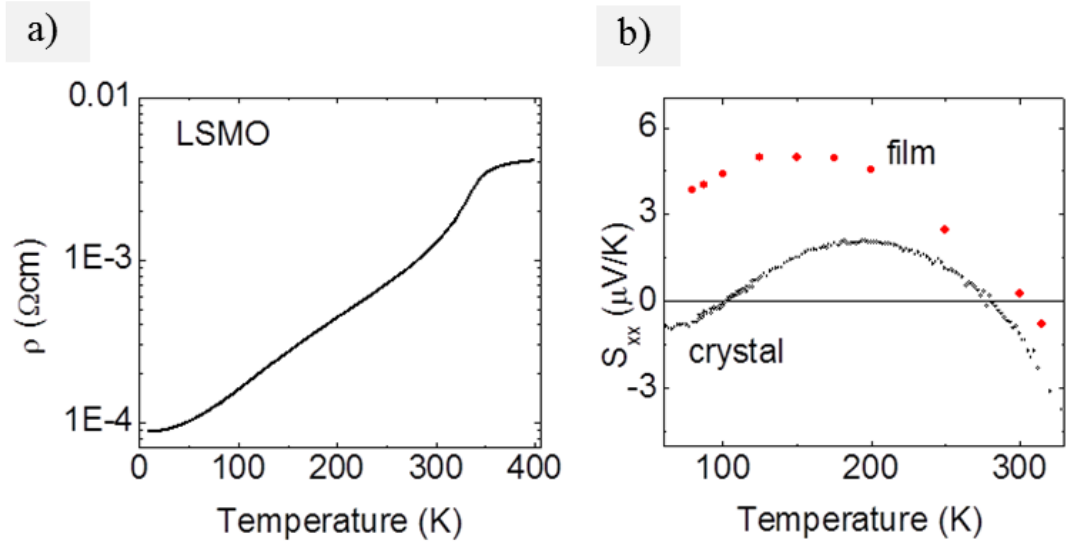

**Figure S5**: Temperature dependence of the electrical resistivity of a LSMO film **(a)**. In **(b)** we show the Seebeck coefficient of a single crystal of LSMO, as well as of the thin-film (35 nm). The Seebeck coefficient in the film was measured using two Pt thermometers, as explained in the text. The solid red dots in the figure represent the raw data, without subtracting the contribution of Pt.